\documentclass[runningheads]{llncs}
\newif\iffinal
\finaltrue

\iffinal
        \newcommand{\karthik}[1]{\textcolor{cyan}{}}
        \newcommand{\jt}[1]{\textcolor{blue}{}}
        \newcommand{\giuliano}[1]{\textcolor{brown}{}}
        \newcommand{\nsh}[1]{\textcolor{green}{}}
        \newcommand{\ag}[1]{\textcolor{purple}{}}
	\else
        \newcommand{\karthik}[1]{\textcolor{cyan}{Karthik: #1}}
        \newcommand{\giuliano}[1]{\textcolor{brown}{Giuliano: #1}}
        \newcommand{\jt}[1]{\textcolor{blue}{Joseph: #1}}
        \newcommand{\nsh}[1]{\textcolor{green}{Shankar: #1}}
        \newcommand{\ag}[1]{\textcolor{purple}{[AG: #1]}}     
\fi

\usepackage{xspace}
\usepackage{listings}
\usepackage{tikz}
\usetikzlibrary{arrows.meta}

\lstdefinestyle{pvs}{
    language={},
    basicstyle=\ttfamily\scriptsize,
    columns=flexible,
    keepspaces=true,
    keywords=[1]{
        theory, end, begin, importing, exporting, assuming, endassuming,
        type, var, const, macro, function, predicate, recursive, measure,
        axiom, theorem, lemma, corollary, obligation, proposition, conjecture, claim, fact,
        and, or, not, implies, iff, exists, forall, lambda,
        if, then, else, endif, cases, of, endcases,
        datatype, constructors, accessor, recognizer,
        table, endtable, cond, endcond,
        library, from, closure,
        auto_rewrite, auto_rewrite_defs, conversion, judgement
    },
    keywords=[2]{
        bool, boolean, nat, naturalnumber, int, integer, real, rational,
        posnat, negint, posint, nonneg_int, nonpos_int,
        posreal, negreal, nonneg_real, nonpos_real,
        string, char, list, array, set, sequence, finite_set,
        tuple, record, union, subtype, type_from, containing,
        below, above, upto, upfrom
    },
    keywords=[3]{
        true, false, TRUE, FALSE, null, empty, emptyset,
        member, subset, union, intersection, difference,
        card, cardinality, finite, infinite,
        min, max, abs, floor, ceiling, mod, rem, div,
        length, append, cons, car, cdr, nth, reverse
    },
    comment=[l]{\%},
    morecomment=[s]{/*}{*/},
    string=[b]",
    morestring=[b]',
    keywordstyle=[1]\color{blue}\bfseries,
    keywordstyle=[2]\color{purple}\bfseries,
    keywordstyle=[3]\color{teal}\bfseries,
    commentstyle=\color{green!60!black}\itshape,
    stringstyle=\color{red},
    numbers=left,
    numberstyle=\tiny\color{gray},
    stepnumber=1,
    numbersep=8pt,
    frame=single,
    frameround=tttt,
    framesep=3pt,
    backgroundcolor=\color{gray!5},
    aboveskip=\medskipamount,
    belowskip=\medskipamount,
    breaklines=true,
    breakatwhitespace=true,
    tabsize=2,
    showtabs=false,
    showspaces=false,
    showstringspaces=false,
    sensitive=false
}

\lstdefinestyle{cex}{
    language={},
    basicstyle=\ttfamily\scriptsize,
    columns=flexible,
    keepspaces=true,
    keywords=[1]{void},
    keywords=[2]{exchangeV10, verify_wheatStays_PATH_PAYMENT_STRICT_SEND,
        applyPriceErrorThresholds, sassert},
    keywords=[3]{},
    comment=[l]{\%},
    morecomment=[s]{/*}{*/},
    string=[b]",
    morestring=[b]',
    keywordstyle=[1]\color{blue}\bfseries,
    keywordstyle=[2]\color{olive}\bfseries,
    keywordstyle=[3]\color{teal}\bfseries,
    commentstyle=\color{green!60!black}\itshape,
    stringstyle=\color{red},
    numbers=left,
    numberstyle=\tiny\color{gray},
    stepnumber=1,
    numbersep=8pt,
    frame=single,
    frameround=tttt,
    framesep=3pt,
    framexleftmargin=3pt,
    backgroundcolor=\color{gray!5},
    aboveskip=\medskipamount,
    belowskip=\medskipamount,
    breaklines=true,
    breakatwhitespace=true,
    tabsize=2,
    showtabs=false,
    showspaces=false,
    showstringspaces=false,
    sensitive=false
}

\newcommand{\prettylstprompt}[0]{
   \lstset{
    basicstyle={\scriptsize\ttfamily}, 
    breaklines=true,breakatwhitespace=true,
}}

\newcommand{\prettylstcpp}[0]{
   \lstset{
    basicstyle={\ttfamily\scriptsize},
    keywordstyle={\color{blue}\bfseries},
    language=C++,
    frameround=fttt,
    breaklines=true,breakatwhitespace=true,
   morekeywords={sassert, assume, 
   RoundingType, Price, 
   int64_t, int32_t, 
   sea\_nd\_i64, sea\_nd\_i32, exchangeV10},
}}

\newcommand{\seahorn}{\textsc{Seahorn}\xspace}
\newcommand{\seabmc}{\textsc{SeaBMC}\xspace}
\newcommand{\sdex}{\texttt{SDEX}\xspace}
\newcommand{\exchange}{\texttt{exchangeV10}\xspace}
\newcommand{\pu}{\textbf{P1}\xspace}
\newcommand{\pd}{\textbf{P2}\xspace}
\newcommand{\pt}{\textbf{P3}\xspace}
\newcommand{\pq}{\textbf{P4}\xspace}

\usepackage[T1]{fontenc}
\usepackage{cite}
\usepackage{amsmath,amssymb,amsfonts}
\usepackage{graphicx}
\graphicspath{ {images/} }
\usepackage{verbatim}
\usepackage{textcomp}
\usepackage{booktabs}
\usepackage{multirow}
\usepackage{syntax}
\usepackage{framed}
\usepackage{url}

\usepackage{semantic}
\usepackage{listings}
\usepackage{graphicx}
\usepackage{subcaption}
\usepackage{booktabs}
\usepackage[llbracket]{stmaryrd}
\usepackage{mathtools}
\usepackage{mathpartir}
\usepackage{tikz} 
\usetikzlibrary{calc, positioning}
\usetikzlibrary{arrows.meta}
\usetikzlibrary{decorations.pathreplacing}
\usepackage{float}
\usepackage{algorithm}
\usepackage{algpseudocode}

\usepackage[pdftex,pagebackref=false]{hyperref} 
\hypersetup{
  colorlinks=true,   
  linkcolor=red,     
  citecolor=green,   
  urlcolor=blue      
}
\usepackage[capitalise]{cleveref}
\usepackage{orcidlink}

\lstdefinestyle{myCppStyle}{
    belowcaptionskip=1\baselineskip,
    breaklines=true,
    frame=single, 
    numbers=left, 
    basicstyle=\footnotesize\ttfamily, 
    keywordstyle=\color{blue}\bfseries,
    commentstyle=\color{purple}\itshape,
    stringstyle=\color{red},
    identifierstyle=\color{black},
    backgroundcolor=\color{gray!10},
    language=C++
}

\begin{document}
\title{Show Me The Money: An Exercise in Proof-Driven Software Understanding}

\author{Joseph Tafese\thanks{Corresponding author.}\inst{1}\orcidlink{0000-0002-4062-0592}, Karthik Nukala\inst{2}, Hassen Sa\"idi\inst{3}\orcidlink{0009-0007-1276-6379}, Natarajan Shankar\inst{2}\orcidlink{0000-0002-8652-8871}, Arie Gurfinkel\inst{1}\orcidlink{0000-0002-5964-6792}, Giuliano Losa\inst{4}\orcidlink{0000-0003-2341-7928}}

\authorrunning{J. Tafese, et al.}
%

\institute{
University of Waterloo, Waterloo, Canada\\
\email{\{jetafese, agurfink\}@uwaterloo.ca}\\
\and
SRI International, Menlo Park, USA \\
\email{\{karthik.nukala, natarajan.shankar\}@sri.com}
\and
Entalus \\
\email{hassen.saidi@entalus.com}
\and
Stellar Development Foundation \\
\email{giuliano@stellar.org}
}

\maketitle              

\begin{abstract}
We present a case study on \textit{proof-driven software understanding} of mature, security-critical infrastructure.
While formal methods are traditionally applied during the design phase, we present our experience applying formal reasoning onto a mature industrial C++ codebase.
We focus on a formal analysis of the core algorithm that implements the Stellar blockchain's \sdex  order book.
By combining large language models (LLMs), Prototype Verification System (PVS), and \seahorn, we are able to prove core properties of the production codebase.
Our approach also identified an inconsistency in documentation related to the reachability of an exception location.
Most importantly, however, we produce artifacts that make it easy for code changes to be checked against established invariants.
This work demonstrates how the strategic combination of theorem proving and model checking provides a path for delivering robust assurance to legacy systems.
\end{abstract}

\section{Introduction}
\label{sec:introduction}

A vast amount of production software delivers safety/value-critical functionality long after its original developers have moved on.
In such systems, correctness does not rest solely on the absence of known bugs but on the preservation of subtle design invariants whose rationale may exist only in developer intuition or informal documentation.
As these systems evolve, the loss of this understanding creates a significant risk since even small changes can violate assumptions that were not explicitly stated, let alone mechanically checked.
We encountered precisely this situation in Stellar’s offer exchange subsystem, \sdex\footnote{\url{https://developers.stellar.org/docs/learn/fundamentals/liquidity-on-stellar-sdex-liquidity-pools}}, that implements the order book at the heart of the Stellar blockchain.
This code has been deployed and exercised in production for over a decade, is carefully documented, and is widely believed to be correct.
Yet its correctness arguments (e.g., rounding behavior, order fairness, etc.) are embedded in comments and complex control flow.
As a result, even minor changes are approached with caution, slowing development, and increasing the cost of assurance.
The challenge is not a lack of testing or care, but the absence of a representation of the intended semantics of the system that can be easily inspected, queried, and evolved.

Formal models are uniquely suited to this role.
They provide a precise vocabulary for stating assumptions, enable invariant-based reasoning, and serve as stable reference points as implementations evolve.
In principle, constructing such a model is the most direct way to recover and preserve the semantic knowledge embedded in mature code.
However, in practice, constructing formal models from existing implementations has been prohibitively difficult.
Recent progress in large language models (LLMs) fundamentally alters this landscape by enabling automatic production of candidate abstractions in formal languages.
By subjecting LLM-generated abstractions to typechecking, proof obligations, and executable analysis, we can validate and refine them into trustworthy models that support real reasoning about the system.

To understand a widely used production system, we analyze the \sdex with an emphasis on \exchange, its core algorithm with $1\,344$ lines of C/C++ code and $758$ lines of comments.
Using the production codebase, we construct an LLM-aided PVS~\cite{cade92-pvs} abstraction with $260$ lines of PVS and $84$ lines of comments.
With this, we prove $16$ lemmas (generated by the LLM) and $22$ proof obligations (generated by the PVS typechecker) to establish key properties of \exchange.
We also encounter an unprovable subgoal which we convert into a \textit{unit proof}~\cite{vvc} (i.e., a symbolic test case) on the C++ code for \seahorn~\cite{SeaHorn} to find a concrete counterexample.
Our experience shows how formal methods can be applied retroactively on mature systems to produce machine-checked artifacts that aid software understanding.

This paper is supported by a public GitHub repository\footnote{\url{https://github.com/jetafese/order-book-core}} including documentation, C++ code, PVS models and proofs, and \seahorn unit proofs.
The rest of the paper is structured as follows:
we present \sdex and the necessary background in~\cref{sec:offer}, the core components of our exercise in~\cref{sec:methodology}, our reflections in~\cref{sec:discussion} and conclude in~\cref{sec:conclusion}.

\section{\sdex: Stellar's Offer Exchange}
\label{sec:offer}

The Stellar network is a blockchain platform that, in addition to smart contracts, supports asset exchange through a built-in central limit order book, commonly called the Stellar Decentralized EXchange (\sdex).
A \emph{central limit order book} is a standard financial mechanism where participants can submit limit orders, also called offers in our context, to buy/sell assets at some maximum/minimum limit price, and all offers are submitted to the same central matching system.
Alongside its native currency -- the Lumen (XLM) -- users may issue custom assets and trade them via this order book.
Users interact with \sdex by submitting transactions, each containing a list of operations to be executed atomically; potential operations include creating new assets, placing offers to exchange assets, and so-called \emph{path payment} operations that allow chaining asset exchanges (e.g., CAD→USD→EUR).

An \emph{offer} to exchange assets specifies (i) an asset to sell, (ii) an asset to buy, (iii) a minimum acceptable price, (iv) a maximum quantity to be sold and (v) an account for sending/receiving assets.
Offers persist in the order book until they are fully executed or canceled.
A \emph{trade} occurs when two offers are matched and assets are transferred between accounts.
Two offers are said to \emph{cross} if their price constraints overlap, meaning that there exists at least one exchange rate acceptable to both parties.
When offers cross, \sdex may execute a trade, subject to constraints and rounding due to the fixed-point representation of quantities.

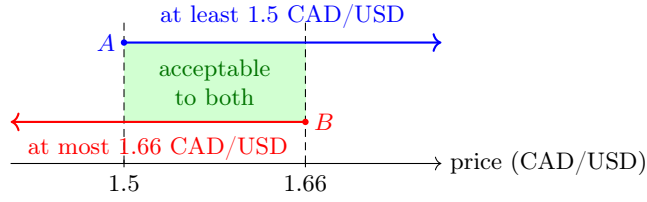
\begin{figure}[t]
\centering
\hspace*{1.8cm}
\begin{tikzpicture}[x=15cm,y=1cm]
\def\xmin{1.40}
\def\xmax{1.78}

\def\yaxis{-1.05}
\def\yA{0.55}
\def\yB{-0.5}
\def\ytop{0.85}
\def\ybot{-0.55}

\fill[green!18] (1.5,\yB) rectangle (1.66,\yA);

\pgfmathsetmacro{\ymid}{(\yA+\yB)/2}
\node[green!45!black,font=\footnotesize,align=center,text width=2.0cm,inner sep=1pt]
at (1.58,\ymid) {acceptable to both};

\draw[->] (\xmin,\yaxis) -- (\xmax,\yaxis) node[right]{price (CAD/USD)};

\draw[densely dashed] (1.5,\yaxis) -- (1.5,\ytop);
\draw[densely dashed] (1.66,\yaxis) -- (1.66,\ytop);
\foreach \x/\lab in {1.5/1.5,1.66/1.66}{
  \draw (\x,\yaxis+0.05) -- (\x,\yaxis-0.05) node[below]{\lab};
}

\draw[thick,blue,->] (1.5,\yA) -- (\xmax,\yA)
node[midway,above=2pt,blue,font=\footnotesize] {at least 1.5 CAD/USD};
\fill[blue] (1.5,\yA) circle (1.2pt);
\node[blue,left] at (1.5,\yA) {$A$};

\draw[thick,red,<-] (\xmin,\yB) -- (1.66,\yB)
node[midway,below=2pt,red,font=\footnotesize] {at most 1.66 CAD/USD};
\fill[red] (1.66,\yB) circle (1.2pt);
\node[red,right] at (1.66,\yB) {$B$};
\end{tikzpicture}
\caption{Crossing offers: $A$ accepts prices $p \ge 1.5$, $B$ accepts prices $p \le 1.66$; the shaded intersection is non-empty.}
\label{fig:crossing}
\end{figure}

We illustrate these concepts with a simple example.
Account $A$ places an offer to sell up to $100$ USD at a price of at least $1.5$ CAD/USD, i.e., it is willing to receive no fewer than $1.5$ CAD for each USD sold.
Account $B$ places an offer to sell up to $200$ CAD at a price of at least $0.6$ USD/CAD.
Expressed in the same units, $B$ requires an exchange rate of at most $1/0.6 \approx 1.66$ CAD/USD and $A$ requires an exchange rate of at least $1.5$ CAD/USD.
Since the acceptable price ranges overlap, the two offers cross.
We show this visually in~\cref{fig:crossing}. 

When two crossing offers are matched, the \sdex computes an exchange that satisfies both parties’ price and quantity constraints after fixed-point rounding.
The exchange executes at most the maximum quantity permitted by both offers.
If the exchange is partial, one offer is fully consumed and removed from the order book, while the other remains with a reduced quantity.
In the example above, one possible execution is to round in $B$'s favor: $A$ sells $100$ USD to $B$ and receives $\lfloor 100 / 0.6 \rfloor = 166$ CAD, with an effective exchange rate of approximately $1.66$ CAD/USD.
However, there are other options, e.g., rounding the other way or picking another valid price.
The exact transfer is determined by \exchange.

The correctness of \sdex is critical to the functioning of the system, as bugs may crash the whole system or lead to the loss of assets and may be exploited by attackers.
Because the \sdex operates on fixed-point arithmetic, its correctness depends on subtle interactions between matching, rounding, and fairness guarantees.
We start with the core algorithm, \exchange, since it is one of the more intricate parts of \sdex.
There are four properties (\pu --- \pq) that must hold at the time of each exchange in order to ensure predictable and economically sound behavior.
\pu: When two offers are matched, rounding is performed in favor of the offer that remains on the order book after the exchange.
\pd: For exchanges that match against a single offer, rounding may alter the effective exchange rate by no more than $1\%$.
\pt: When a transaction is a path payment, rounding errors may exceed $1\%$; for all other transactions, any exchange that would violate this bound is rejected.
\pq: No unexpected exceptions are thrown.

\exchange is primarily implemented in a single file\footnote{See \href{https://github.com/stellar/stellar-core/blob/1d2f58f5a1f47c96978ef507502943174258135c/src/transactions/OfferExchange.cpp}{\texttt{OfferExchange.cpp}}} in the \texttt{stellar-core} repository.
The inputs are: a price, a rounding type, and limit amounts for both sides of a trade, and the output is the amounts exchanged.
We summarize \exchange and (later) its helper functions in Algorithms~\ref{alg:exchangeV10} and~\ref{alg:priceThresholds} respectively.
The function was developed and maintained for $11$ years by at least $5$ primary engineers with a nuanced understanding of the functional behavior and corresponding invariants.
However, organizational turnover made it hard to maintain knowledge of this codebase.

\begin{algorithm}[t]
\caption{\exchange}
\label{alg:exchangeV10}
\begin{algorithmic}[1]
\Require
Price $(n,d)$,
bounds $(W_s^{\max}, W_r^{\max}, S_s^{\max}, S_r^{\max})$,
rounding mode $\mathit{round}$
\Ensure
$(W_r, S_s, \mathit{wheatStays})$

\State $(W_r, S_s, \mathit{wheatStays}) \gets
\Call{ExchangeCore}{(n,d), W_s^{\max}, W_r^{\max}, S_s^{\max}, S_r^{\max}, \mathit{round}}$

\State $(W_r, S_s) \gets
\Call{ApplyPriceErrorThresholds}{(n,d), W_r, S_s, \mathit{wheatStays}, \mathit{round}}$

\State \Return $(W_r, S_s, \mathit{wheatStays})$
\end{algorithmic}
\end{algorithm}

\section{Proof-Driven Software Understanding}
\label{sec:methodology}

\begin{figure}[t]
\centering
\includegraphics[width=\textwidth, alt={The three stages of our exercise: Formalizing, Proving and Finding Defects in Industrial C++ Code using LLMs, PVS and SeaHorn}]
{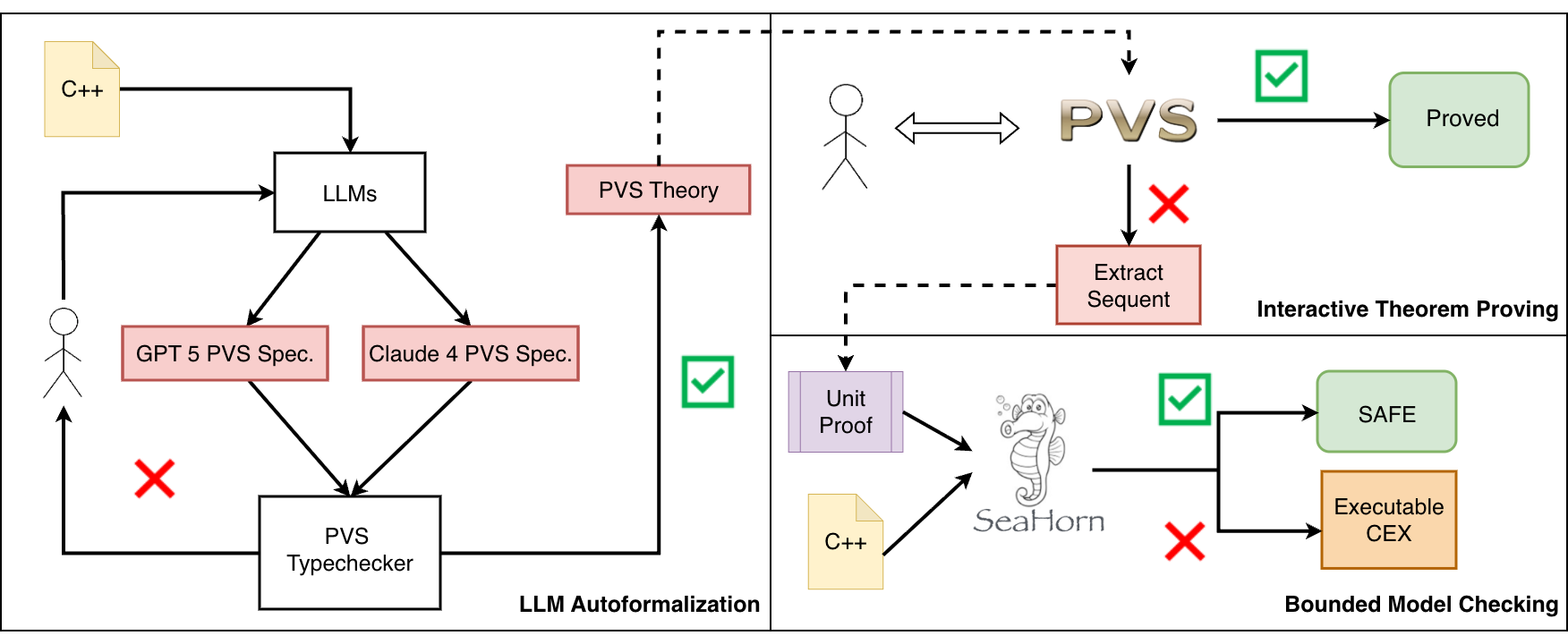}
\caption{We present the three stages of our exercise: Formalizing, Proving and Finding Defects in Industrial C++ Code using LLMs, PVS and SeaHorn.}
\label{fig:arch}
\end{figure}

To understand the \exchange implementation and produce a machine checkable artifact, we aim at:
(1) constructing a formal model that captures its core functionality, 
(2) proving that properties (\pu{} -- \pq) hold of the model and
(3) ensuring they also hold for the actual code.
We present a streamlined approach (\cref{fig:arch}) that 
(1) leverages LLMs to autoformalize the \exchange code with feedback from the PVS typechecker (\cref{subsec:cpp2pvs}), 
(2) proves properties of \exchange aided by PVS's automation (\cref{sec:pvs-iter}), and
(3) verifies that the model corresponds to the original C++ code using \seabmc~\cite{seabmc}, the bounded model checker of \seahorn (Section \ref{subsec:cex}).

\subsection{LLM Autoformalization (C++ to PVS)}
\label{subsec:cpp2pvs}

At this stage, our goal was to go from a blank screen to a potentially useful PVS model that successfully passes the PVS typechecker, a first anchor for understanding the codebase.
We leveraged the fact that the PVS language is functional and \exchange is mostly functional, in that it does not maintain state but can throw exceptions.
\exchange contains both code and natural-language arguments of established properties and invariants, expressed as comments accompanying the C++ code.
The goal is to produce a formalization in PVS that is faithful to the code and comments (see~\cref{fig:autoformalization-example}).

\begin{figure}[t]
\centering
\begin{lstlisting}[style=pvs, numbers=none]
%---------------------------------------------------------------------------
% less than or equal to 1% bound when cFW is False. That is:
% abs((price`n/price`d)-(sS/wR))/((price`n/price`d)) <= 1/100
%---------------------------------------------------------------------------
one_percent_normal: LEMMA
FORALL (p: Price, wR: posint64, sS: posint64):
  LET price = (p`n/p`d), effP = (sS/wR) IN
  checkPriceErrorBound(p, wR, sS, False) IFF abs(price - effP)/(price) <= 1/100
%---------------------------------------------------------------------------
% less than or equal to 1% bound when cFW is True and sheep is favored.
% That is: abs((price`n/price`d)-(sS/wR))/((price`n/price`d)) <= 1/100
%---------------------------------------------------------------------------
one_percent_path_favor_sheep: LEMMA
FORALL (p: Price, wR: posint64, sS: posint64):
  LET price = (p`n/p`d), effP = (sS/wR), rhs = 100 * p`d * sS, lhs = 100 * p`n * wR IN
  checkPriceErrorBound(p, wR, sS, True) AND (rhs < lhs) => abs(price - effP)/(price) <= 1/100
\end{lstlisting}
\caption{Autoformalized lemma from the comment: ``Check that the relative error between the price and the effective price does not exceed 1\%. If canFavorWheat == true then this function does an asymmetric check such that error favoring the seller of wheat can be unbounded, while the relative error between the price and the effective price does not exceed 1\% if it is favoring the seller of sheep. The functionality of canFavorWheat (cFW) is required for \texttt{PathPayment}.''}
\label{fig:autoformalization-example}
\end{figure}

We started by copying the full source code of \exchange to ChatGPT 5 Pro, prompting it to generate a PVS model, and validating the model with a local PVS client.
The PVS typechecker had several complaints that we resolved by copying failure messages to ChatGPT and copying the PVS model it produces back to our PVS client.
We repeated the process with Claude Opus 4, to similar results.
At this stage, we encountered hallucinations such as \texttt{x DIV y} (where \texttt{DIV} is not a PVS function), and declaration-level syntactic issues like misplaced \texttt{IMPORTING} statements and made-up library names.
Once we addressed these typechecking failures\footnote{PVS's full typechecking involves the generation of TCCs, rendering type inhabitation undecidable in general. We refer, here, to the first phase which involves name resolution and sound generation of TCCs.}, we found that there was significant structural correspondence between \exchange and the PVS model.
For example, function boundaries were respected, conditional statements were in direct correspondence and the generated lemmas aligned with the comments in \exchange.
On the other hand, the LLMs wrongly modeled exceptions as unused boolean variables, which hindered attempts to prove \pq.
To capture the learning from this exercise, we developed a prompt (see~\cref{fig:cpp2pvs-prompt}) that successfully reduced these pitfalls.

Through interaction with PVS, we manually separated the theories that correspond to distinct functions used by \exchange into $3$ separate files (i.e., a file for \exchange, \texttt{ApplyPriceErrorThresholds} and \texttt{CheckPriceErrorBound}).
This allowed us to isolate properties by their relevance to a function and define preconditions to determine how functions should interface with each other.
The latter capability proved to be quite a powerful way to specify exception-free paths through a function (i.e., the  ``happy path'').

The rapid iteration with the LLM allowed us to understand the software being formalized without spending significant manual effort.
While there are many opportunities for automating the otherwise manual parts of this stage, we explicitly choose to leave the human-in-the-loop because our exercise centers around \textit{human} software understanding.
Moreover, this exercise can serve as a gentle introduction to software modeling and theorem proving.

\subsection{Theorem Proving and Efficient Arguments in PVS}
\label{sec:pvs-iter}

\begin{figure}[t]
\centering
\begin{subfigure}[b]{0.5\textwidth}
\begin{lstlisting}[style=pvs, numbers=none]
one_percent_path_favor_sheep: LEMMA
FORALL (p: Price, wR: posint64, sS: posint64):
LET price = (p`n/p`d), effP = (sS/wR),
    rhs = 100 * p`d * sS, lhs = 100 * p`n * wR
IN
  checkPriceErrorBound(p, wR, sS, True)
  AND (rhs < lhs) 
  IMPLIES abs(price - effP)/(price) <= 1/100
\end{lstlisting}
\end{subfigure}%
\hfill%
\begin{subfigure}[b]{0.43\textwidth}
\begin{lstlisting}[style=pvs, numbers=none]
% 1) Skolemization, exposing predicates
(SKEEP :PREDS? T)
% 2) Shostak/SMT simplification
(ASSERT) ; Shostak
% 3) BDD simplifier/normalizer
(BDDSIMP)
% 4) Super-duper rewrite/expand/SMT/BDD
(GRIND :THEORIES ("real_props"))
\end{lstlisting}
\end{subfigure}
\caption{An example PVS proof of an asymmetrical rounding lemma from~\cref{fig:autoformalization-example}.}
\label{fig:dp}
\end{figure}

\begin{figure}[t]
    \centering
\begin{minipage}[t]{\textwidth}
\prettylstcpp
\begin{lstlisting}
// --[OfferExchange.cpp, L164-166]--
// These never overflow since price.n and price.d are int32_t
int64_t errN = (int64_t)100 * (int64_t)price.n;
int64_t errD = (int64_t)100 * (int64_t)price.d;
\end{lstlisting}
\end{minipage}
\noindent
\begin{minipage}[t]{0.5\textwidth}
\begin{lstlisting}[style=pvs, numbers=none]
Price: TYPE = [# n: uint32, d: uint32 #]
checkPriceErrorBound(price: Price...): bool 
= LET errN: posint64 = 100 * price`n,
      errD: posint64 = 100 * price`d,
      ...
\end{lstlisting}
\begin{lstlisting}[style=pvs, numbers=none]
----[Prelude integertypes]----
uint32: TYPE = upto(exp2(32) - 1)
uint64: TYPE = upto(exp2(64) - 1)
posint64: TYPE = { u: uint64 | u > 0 }
times_u16_u16: JUDGEMENT
  *(ux16, uy16) HAS_TYPE uint32
times_u32_u32: JUDGEMENT
  *(ux32, uy32) HAS_TYPE uint64
\end{lstlisting}
\end{minipage}%
\hfill%
\begin{minipage}[t]{0.45\textwidth}
\vspace{1.8em} 
\begin{lstlisting}[style=pvs, numbers=none]
% Subtype TCC generated for  100 * price`n
% expected type  posint64
checkPriceErrorBound_TCC1: OBLIGATION
  FORALL (price: Price, ...):
    100 * price`n > 0;

% Subtype TCC generated for 100 * price`d
% expected type  posint64
checkPriceErrorBound_TCC2: OBLIGATION
  FORALL (price: Price, ...):
  100 * price`d > 0;
\end{lstlisting}
\end{minipage}
\caption{We examine an \exchange lemma derived from comments in \texttt{OfferExchange.cpp},shown at the top.
During autoformalization, these declarations become \texttt{LET}-bindings as shown in the top-left \texttt{checkPriceErrorBound} PVS declaration.
In the bottom-left, we show that leveraging the existing PVS Prelude theory, the typechecker can automatically establish type constraints across type families.
On the right, we show type-correctness conditions (TCCs) generated by the typechecker that need to be discharged in the PVS prover.}
\label{fig:tccs}
\end{figure}

PVS's emphasis on the interplay between language and automation facilitates both the autoformalization effort and the interactive human proof effort.
The PVS language supports a style of formalization that restricts properties to a minimal scope, ensuring tractability and predictability of the underlying decision procedures (involved in discharging these conditions) and providing just enough feedback so the human can understand what might needed for correctness.

To prove the properties of \exchange, we use PVS's automated internal procedures (e.g., \texttt{assert}, \texttt{grind}) built on to Shostak's method~\cite{Shostak:combination}.
We also discharge queries involving nonlinear arithmetic with the Yices2 SMT solver \cite{Dutertre:cav2014}.
Shostak's method enables incremental SMT reasoning and proof state simplification, allowing a user-facing entry point to the internal solver and tractable readings of its current solution state.
\cref{fig:dp} showcases the variety of procedures used in establishing the asymmetric rounding condition of the \exchange.
In addition to this lemma, there were typechecking constraints that needed to be established to complete the proof.
We show examples of these in~\cref{fig:tccs}.

In this stage, we gained local, property-directed understanding of the program by leveraging PVS's modular language and representations.
When combined with transparent automation and specification utilities, program understanding became a dialogue with the proof assistant.
We proved $37/38$ lemmas and TCCs that were sufficient to prove correctness of \exchange relative to \pu --- \pt.
For example, for \pd, we proved that the PVS model adhered strictly to the $1\%$ bound on rounding for exchanges that match a single offer.
On the $38$'th TCC, however, we were confronted with a subgoal for \pq that we could not prove (after numerous attempts and iterations); this prompted us to go back to the C++ code.

\subsection{SeaHorn and Executable Counter Examples}
\label{subsec:cex}

We found it useful to connect our inability to prove \pq in PVS with a counterexample execution of the C++ code.
Inspired by the Floyd-Hoare Logic~\cite{floyd1993, hoare1969}, and much like a unit test, a unit proof consists of a harness that sets up symbolic variables and preconditions, invokes the system under test (SUT), and asserts postconditions.
We constructed unit proofs by encoding the type constraints from the PVS formalization as assumptions over C++ variables, and composed assertions using standard Boolean operators.
\seahorn compiles the unit proof and the SUT to LLVM bitcode, which is then checked by the \seabmc model-checker as demonstrated in~\cite{simulation, tafese2025tale, btor2mlir, dissertation}.
For example, we used the executable witness (read off the PVS sequent/unprovable subgoal) to construct the unit proof in~\cref{fig:unit-proof}, from which \seahorn generated an executable counterexample (CEX) that we debugged with \texttt{lldb} (see the trace in~\cref{fig:cex}).
This is how we disproved a property in the comments about a path-payment exception that is not expected to be thrown.
Concretely, \pq is disproved since an exception is thrown when the rounding type is a Path Payment Strict Send and the inputs to \exchange are: $p=(2, 4), \mathit{maxWheatSend} = 1, \mathit{maxWheatReceive} = 2, \mathit{maxSheepSend} = 1, \mathit{maxSheepReceive} = 0$.
We present a pseudo-code description of the problematic function in~\cref{alg:priceThresholds}, where the exception condition in question appears as an assert at line 15.
\sdex developers are confident that upstream code prevents this exception from being reached, but proving this claim is out-of-scope for \exchange.

\newsavebox{\figpathunitproof}
\begin{lrbox}{\figpathunitproof}
\prettylstcpp
\begin{lstlisting}[breaklines=true,numbers=right]
void verify_wheatStays_PATH_PAYMENT_STRICT_SEND() {
    // setup variables provided by called
    int32_t n = sea_nd_i32();
    int32_t d = sea_nd_i32();
    int64_t maxWheatSend = sea_nd_i64();
    int64_t maxSheepReceive = sea_nd_i64();
    int64_t maxWheatReceive = sea_nd_i64();
    int64_t maxSheepSend = sea_nd_i64();
    // assume preconditions
    assume(n > 0 && d > 0);
    assume(maxWheatSend >= 0 && maxSheepSend > 0);
    assume(maxWheatReceive > 0);
    auto round = RoundingType::PATH_PAYMENT_STRICT_SEND;
    // Call SUT
    Price p{n, d};
    auto res = exchangeV10(p, maxWheatSend, maxWheatReceive, maxSheepSend,
        maxSheepReceive, round);
    // enforce postconditions
    sassert(!res.wheatStays || (res.numSheepSend > 0 && 
        res.numWheatReceived >= 0));
}
\end{lstlisting}
\end{lrbox}%
\begin{figure*}[t]
\scalebox{0.95}{\usebox{\figpathunitproof}}
\caption{Unit proof for \exchange with Path Payment Strict Send Rounding.}
\label{fig:unit-proof}
\end{figure*}

Integrating \seahorn complements the proof-based methodology in several ways. 
First, it acts as a sanity check, ensuring that formally proven invariants are relevant and observable in the concrete system. Once a lemma was proven in PVS, we replicated it with a unit proof, over the \exchange implementation, to be run with \seahorn. 
Second, it accelerates debugging by producing concrete inputs that trigger subtle corner cases.
Finally, it enables a tightly coupled iterative loop: PVS abstractions guide the generation of unit proof preconditions, counterexamples inform refinement of the abstraction, and updated proofs confirm the corrected properties.
This provides a pathway for developers to gain confidence in the correctness of legacy code while still supporting code evolution.

\begin{algorithm}[ht]
\caption{Apply Price Error Thresholds}
\label{alg:priceThresholds}
\begin{algorithmic}[1]
\Require
Price $(n,d)$, amounts $(W_r, S_s)$, $\mathit{wheatStays}$, rounding mode
\Ensure Adjusted $(W_r, S_s)$

\Function{ApplyPriceErrorThresholds}{$(n,d), \mathit{maxSend}, \mathit{maxReceive}$}
\If{$W_r > 0 \land S_s > 0$}
    \State $V_r \gets W_r \cdot n$
    \State $V_s \gets S_s \cdot d$
    \State \textbf{assert:} $\mathit{wheatStays} \implies V_s \geq V_r$  \textbf{and} $\neg \mathit{wheatStays} \implies V_s \leq V_r$

    \If{$\mathit{round} = \textsf{NORMAL}$}
        \If{$\neg$ \Call{CheckPriceErrorBound}{$(n,d), W_r, S_s, false$}}
            \State $W_r \gets 0;\; S_s \gets 0$
        \EndIf
    \Else
        \State \textbf{assert:} \Call{CheckPriceErrorBound}{$(n,d), W_r, S_s, true$}
    \EndIf
\Else
    \State \textbf{assert:} $\mathit{round} = \textsf{PPS\_SEND} \implies S_s \neq 0$
    \State $W_r \gets 0;\; S_s \gets 0$
\EndIf
\State \Return $(W_r, S_s)$
\EndFunction

\Function{CheckPriceErrorBound}{$(n,d), (W_r, S_s), \mathit{canFavourWheat}$}
\State $K \gets 100$ \Comment{Threshold scaling factor for 1\% error}
\State $L \gets K \cdot n \cdot W_r; \; R \gets K \cdot d \cdot S_s$
\If{$\mathit{canFavorWheat} \land R > L$}
\State \Return \textbf{true}
\EndIf
\State $D \gets |L - R|$ \Comment{Absolute difference between scaled values}
\State $C \gets n \cdot W_r$ \Comment{Cap based on intended price}
\State \Return $(D \le C)$
\EndFunction
\end{algorithmic}
\end{algorithm}

\section{Discussion}
\label{sec:discussion}

We have presented our experience formally specifying an industrial C++ code in PVS where we successfully proved properties \pu{}---\pt and disproved \pq.
While these properties were so far only informally documented, we now have a validated model and proofs about the implementation. 
In this section we discuss the broader applicability of the methodology and areas of improvements and future research.

Our success in formalizing and validating the model of \exchange is in part due to the fact that the C++ code involved is well-documented, including informal explanations of critical invariants and their proofs, mostly functional, and modular in its structure and presentation.
The informal documentation and invariants helped us and the LLM formalize the properties to prove.
Without this, we would have had to learn a lot more about the subsystem and the context in which it is used in order to derive the important properties to check.
Modularity of the code allowed us to proceed step by step, without any blow-up in complexity as we formalized more and more code. 
By following good engineering principles, the original authors of this code already did a good part of the work involved in applying formal methods to gain assurance in software correctness, and the methodology we present allows harnessing that work.

\begin{figure}[t]
\begin{lstlisting}[style=cex, numbers=none]
(lldb) target create "./exchange.out"
Current executable set to '/order-book-core/exchange/exchange.out' (arm64).
(lldb) b -n __VERIFIER_error                                                                               
Breakpoint 1: where = exchange.out`::__VERIFIER_error() + 8 at seahorn.cpp:26:3, ...
(lldb) r
Process 71139 launched: '/order-book-core/exchange/exchange.out' (arm64)
Process 71139 stopped
* thread #1, queue = 'com.apple.main-thread', stop reason = breakpoint 1.1
    frame #0: 0x0000000100012824 exec.out`::__VERIFIER_error() at seahorn.cpp:26:3 [opt]
   23   }
   24  
   25   void __VERIFIER_error() {
-> 26     printf("[sea] __VERIFIER_error was executed\n");
   27     exit(1);
   28   }
   29  
Target 0: (exec.out) stopped.
(lldb) bt
* thread #1, queue = 'com.apple.main-thread', stop reason = breakpoint 1.1
  * frame #0: 0x0000000100012824 exec.out`::__VERIFIER_error() at seahorn.cpp:26:3 [opt]
    frame #1: 0x0000000100000e50 exec.out`sassert(cond=false) at vOfferExchange.cpp:13:9
    frame #2: 0x0000000100002694 exec.out`stellar::applyPriceErrorThresholds(price=(n = 2, d = 4), wheatReceive=0, sheepSend=0, wheatStays=false, round=PATH_PAYMENT_STRICT_SEND) at vOfferExchange.cpp:790:17
    frame #3: 0x000000010000213c exec.out`stellar::exchangeV10(price=(n = 2, d = 4), maxWheatSend=1, maxWheatReceive=2, maxSheepSend=1, maxSheepReceive=0, round=PATH_PAYMENT_STRICT_SEND) at vOfferExchange.cpp:584:12
    frame #4: 0x0000000100000814 exec.out`verify_wheatStays_PATH_PAYMENT_STRICT_SEND() at verify.cpp:54:16
    frame #5: 0x0000000100000738 exec.out`main at verify.cpp:25:5
    frame #6: 0x000000018746eb98 dyld`start + 6076
\end{lstlisting}
\caption{Debugging the \seahorn Generated CEX with \texttt{lldb}.}
\label{fig:cex}
\end{figure}

We expect that our methodology will apply equally well to other code bases that follow strong engineering principles and are well-documented and modular, even if they are larger.
The key reason is that the formal artifacts we construct (e.g., proof-oriented abstractions, invariants, and lemmas) are aligned with the code structure and subsystem boundaries, inheriting the modularity and separation of concerns introduced by the engineers who authored the code.
One important aspect that we have not experienced in practice is the adaptation to an evolving code base (\exchange did not change during our exercise).
The \seahorn unit proofs are likely to be of help in this case because they can identify property violations as long as the interfaces used by the unit proofs do not change.
Otherwise, we have to reapply the autoformalization methodology and create new unit proof harnesses.
Automating the creation of \seahorn harnesses would help in this process, and this is an area where our work can be improved.

An important outcome of this work is that it provides a concrete narrative that formal methods researchers can use when engaging with industry.
Rather than requiring greenfield development or specialized training, we start from artifacts engineers already produce: production code, comments, informal invariants, and test harnesses.
LLM-assisted autoformalization lowers the barrier to constructing an initial model; PVS provides strong automation and early feedback; and \seahorn reconnects formal reasoning to executable behavior.
This combination makes it feasible to integrate formal reasoning into existing workflows such as code review, regression testing, and continuous integration.
In the \exchange codebase, engineers have reasoned explicitly about rounding, fairness, bounds, and unreachable states, but these arguments are scattered across comments and control flow.
Our approach captures this effort in formal, machine-checked proofs, reducing the risk of knowledge loss and making future changes less error-prone.
In this sense, formal methods act as a force multiplier on existing diligence rather than a replacement for it.

From a research perspective, this case study exposes several concrete directions that merit further investigation.
First, the translation from deductive proof artifacts (such as PVS sequents) to model checking harnesses is highly structured and ripe for automation.
Automating this step would strengthen the connection between abstract proofs and concrete executions, enabling tighter feedback loops.
Second, richer support for model finding and counterexample generation at the specification level would help identify underspecification earlier, reducing downstream proof effort.
This can be achieved by debugging an ill-specified lemma using PVS's integrated type-directed random tester~\cite{Owre2006} and the PVSio ground evaluation framework~\cite{Mun03NASA}.
Third, improving the reliability, validation, and explainability of LLM-assisted formalization remains an open challenge with clear industrial impact.
Finally, given proof artifacts (e.g., PVS formalizations, \seahorn unit proofs) for a codebase, maintaining them in light of evolving specifications is highly valuable.
For example, progress in generating and deploying code changes based on updated proof artifacts ensures only proven contributions make it to production.
On the other hand, enabling the automatic updating of proof artifacts given code changes ensures that understanding is not lost as a codebase evolves.

\begin{figure}[t]
\prettylstprompt
\begin{lstlisting}[breaklines=true,numbers=left]
<role>
You are an expert in formal methods, program semantics, and the Prototype
Verification System (PVS). Your task is to translate C++ code into a
proof-oriented PVS specification that captures the intended mathematical
behavior of the algorithm.
</role>
<instructions>
1. Read the C++ code carefully to understand the algorithmic intent,
   control flow, and arithmetic structure.
2. Ignore C++-specific operational details such as memory management,
   logging, profiling, and performance optimizations.
3. Translate the code into a PVS theory that:
   (a) uses total functions,
   (b) replaces exceptions and runtime errors with explicit preconditions
       or predicate subtypes,
   (c) models arithmetic over mathematical integers with explicit bounds.
4. Preserve the functional behavior of the algorithm under its intended
   assumptions, but do NOT attempt to model full C++ semantics.
5. For each inline comment or "should never happen" assertion in the C++
   code, introduce either:
   (a) a named lemma stating the assumed invariant, or
   (b) a strengthened precondition on the corresponding PVS definition.
6. Use idiomatic PVS syntax and structure the specification to facilitate
   automated reasoning.
7. Do not attempt to prove any lemmas; only state them precisely.
8. Clearly separate: (a) core computational definitions, (b) policy decisions (e.g., rounding modes), (c) assumed invariants.
</instructions>
<example> <inputs> <cpp_code> ... </cpp_code> </inputs>
<output> <pvs_spec> ... </pvs_spec> </output> </example>
Now generate the output for the following input. The output must be a
well-formed PVS theory enclosed in <output></output> tags.
<inputs> <cpp_code> ${cpp_code} </cpp_code> </inputs>
\end{lstlisting}
\label{fig:cpp2pvs-prompt}
\caption{LLM Prompt Template for C++ to PVS Abstraction.}
\end{figure}

Taken together, these observations suggest a complementary role for formal methods in industrial software development.
Rather than insisting on full formalization upfront, our approach supports incremental adoption, retrospective analysis, and continuous validation.
This aligns naturally with how critical systems are actually built and maintained, and it opens the door to new research that is directly informed by industrial constraints.
We believe this perspective is essential for broadening the impact of formal methods and for ensuring their relevance to the next generation of critical and long-lived software systems.

\section{Conclusion}
\label{sec:conclusion}

We have demonstrated that PVS-based abstraction engineering and bounded model checking with \seahorn can enable safe, iterative evolution of legacy, high-value blockchain infrastructure.
By combining mechanized proof and symbolic validation, we converted semi-formal, comment-based reasoning into a verifiable and extensible correctness pipeline. 
Our ongoing efforts will finalize the specification and expand property coverage to additional components of \sdex.

\section{Data-Availability Statement}

The data supporting the results reported in your paper can be accessed at the following link: \url{https://doi.org/10.5281/zenodo.20125587}.
We have included an artifact in a docker environment that captures the C/C++ code, \seahorn proofs, PVS proofs and their respective environments.

%
%
%
\bibliographystyle{splncs04}
\bibliography{bibliography}

\end{document}